\pdfoutput=1

\documentclass[sigconf]{acmart}
\AtBeginDocument{%
  \providecommand\BibTeX{{%
    \normalfont B\kern-0.5em{\scshape i\kern-0.25em b}\kern-0.8em\TeX}}}

\copyrightyear{2020} 
\acmYear{2020} 
\setcopyright{acmcopyright}
\acmConference[IH\&MMSec '20] {2020 ACM Workshop on Information Hiding and Multimedia Security}{June 22--24, 2020}{Denver, CO, USA}
\acmBooktitle{2020 ACM Workshop on Information Hiding and Multimedia Security (IH\&MMSec'20), June 22--24, 2020, Denver, CO, USA}
\acmPrice{15.00}
\acmDOI{10.1145/3369412.3395063}
\acmISBN{978-1-4503-7050-9/20/06} 

\acmSubmissionID{34}
\usepackage{amsmath,amsfonts,amssymb,psfrag}
\usepackage{amsthm}
\usepackage{graphicx}
\usepackage{epstopdf}
\usepackage{rotating}
\usepackage{bm}
\usepackage{dsfont}
\usepackage{color}
\usepackage{tikz}
\usetikzlibrary{plotmarks}
\usetikzlibrary{shapes,arrows,fit,calc,positioning,automata}
\usepackage{pgfplots}
\usetikzlibrary{calc}
\usetikzlibrary{shapes,arrows}
\usetikzlibrary{decorations.markings}
\usetikzlibrary{positioning}
\pgfplotsset{compat=1.10}
\usetikzlibrary{calc}
\usetikzlibrary{shapes,arrows}
\usetikzlibrary{decorations.markings}
\usepackage{mathtools}
\usepackage{verbatim, hyperref}

\newcommand{\vq}{\text{q}}
\newcommand{\fec}{\text{s}}
\newcommand{\quantCode}{\ensuremath{\mathcal{C}_{\vq} }}
\newcommand{\fecCode}{\ensuremath{\mathcal{C}_{\fec} }}
\newcommand{\quantR}{\ensuremath{R_{\vq} }}
\newcommand{\fecR}{\ensuremath{R_{\fec} }}
\newcommand{\quantDim}{\ensuremath{K_{\vq} }}
\newcommand{\fecDim}{\ensuremath{K_{\fec} }}
\newcommand{\quantk}{\ensuremath{k_{\vq} }}
\newcommand{\feck}{\ensuremath{k_{\fec} }}

\newcommand{\eqdef}{\ensuremath{\stackrel{\mathclap{\mbox{\normalfont\small def}}}{=}}}

\newcommand{\pA}{\ensuremath{p_{\text{A}}}}
\newcommand{\pc}{\ensuremath{p_{\text{c}}}}
\newcommand{\pB}{\ensuremath{P_{\text{B}}}}

\newcommand{\dmin}{\ensuremath{d_{\text{min}}}}
\newcommand{\dfree}{\ensuremath{d_{\text{free}}}}
\newcommand{\Afree}{\ensuremath{A_{\text{free}}}}
\newcommand{\UB}{\ensuremath{P_{B}^{\text{UB}}}}

\newcommand{\ccenc}[3]{\ensuremath{[#1,#2,#3]}}

\newcommand{\assignRand}{\ensuremath{\xleftarrow{\$}}}

\newcommand{\F}[1]{\ensuremath{\mathbb{F}_{#1}}}
\newcommand{\Fq}{\F{2}}

\newcommand{\Enc}{\mathsf{Enc}}
\newcommand{\Dec}{\mathsf{Dec}}

\usepackage[ruled,vlined,titlenumbered,linesnumbered]{algorithm2e}

\begin{document}
\fancyhead{}
\title{Nested Tailbiting Convolutional Codes for \\Secrecy, Privacy, and Storage}

\author{Thomas Jerkovits}
\email{thomas.jerkovits@dlr.de}
\affiliation{%
  \institution{German Aerospace Center}
  \city{Weßling}
  \state{Germany}
}

\author{Onur G\"unl\"u}
\email{guenlue@tu-berlin.de}
\affiliation{%
  \institution{TU Berlin}
  \city{Berlin}
  \country{Germany}}

\author{Vladimir Sidorenko}
\email{vladimir.sidorenko@tum.de}
\author{Gerhard Kramer}
\email{gerhard.kramer@tum.de}
\affiliation{%
  \institution{TU Munich}
  \city{Munich}
  \country{Germany}
}


\begin{abstract}
A key agreement problem is considered that has a biometric or physical identifier, a terminal for key enrollment, and a terminal for reconstruction. A nested convolutional code design is proposed that performs vector quantization during enrollment and error control during reconstruction. Physical identifiers with small bit error probability illustrate the gains of the design. One variant of the nested convolutional codes improves on the best known key vs. storage rate ratio but it has high complexity. A second variant with lower complexity performs similar to nested polar codes. The results suggest that the choice of code for key agreement with identifiers depends primarily on the complexity constraint.
\end{abstract}

\ccsdesc[500]{Security and privacy~Information-theoretic techniques}

\keywords{nested codes, information privacy, tailbiting, convolutional codes, physical unclonable functions}

\maketitle

\section{Introduction}
Irises and fingerprints are biometric identifiers used to authenticate and identify individuals, and to generate secret keys \cite{Campisi}. In a digital device, there are digital circuits that have outputs unique to the device.
One can generate secret keys from such physical unclonable functions (PUFs) by using their outputs as a source of randomness.
Fine variations of ring oscillator (RO) outputs, the start-up behavior of static random access memories (SRAM), and quantum-physical readouts through coherent scattering \cite{QRPUF} can serve as PUFs that have reliable outputs and high entropy \cite{IgnaCTW,GassendThesis}. One can consider them as physical ``one-way functions'' that are easy to compute and difficult to invert \cite{PappuThesis}. 

There are several security, privacy, storage, and complexity constraints that a PUF-based key agreement method should fulfill. First, the method should not leak information about the secret key (negligible \textit{secrecy leakage}).
Second, the method should leak as little information about the identifier (minimum \textit{privacy leakage}). 
The privacy leakage constraint can be considered as an upper bound on the secrecy leakage via the public information of the first enrollment of a PUF about the secret key generated by the second enrollment of the same PUF \cite{benimdissertation}. 
Third, one should limit the \textit{storage} rate because storage can be expensive and limited, e.g., for internet-of-things (IoT) device applications. Similarly, the hardware cost, e.g., hardware area, of the encoder and decoder used for key agreement with PUFs should be small for such applications. 

There are two common models for key agreement: the \textit{generated-secret (GS)} and the \textit{chosen-secret (CS) models}. 
An encoder extracts a secret key from an identifier measurement for the GS model, while for the CS model a secret key that is independent of the identifier measurements is given to the encoder by a trusted entity. 
In the classic key-agreement model introduced in \cite{AhlswedeCsiz} and \cite{Maurer}, two terminals observe correlated random variables and have access to a public, authenticated, and one-way communication link; an eavesdropper observes only the public messages called \textit{helper data}. The regions of achievable secret-key vs. privacy-leakage (key-leakage) rates for the GS and CS models are given in \cite{IgnaTrans,LaiTrans}.
The storage rates for general (non-negligible) secrecy-leakage levels are analyzed in \cite{storage}, while the rate regions with multiple encoder and decoder measurements of a hidden source are treated in \cite{bizimTIFSMultipleMeasurement}.
There are other key-agreement models with an eavesdropper that has access to a sequence correlated with the identifier outputs, e.g., in \cite{csiszarnarayan,Khisti,Blochpaper,benimdissertation}. This model is not realistic for PUFs, unlike physical-layer security primitives and some biometric identifiers that are continuously available for physical attacks. PUFs are used for \textit{on-demand} key reconstruction, i.e., the attack should be performed during execution, and an invasive attack applied to obtain a correlated sequence permanently changes the identifier output \cite{GassendThesis,bizimpaper}.
Therefore, we assume that the eavesdropper cannot obtain a sequence correlated with the PUF outputs. 

Two classic code constructions for key agreement are code-offset fuzzy extractors (COFE) \cite{Dodis2008fuzzy} and the fuzzy commitment scheme (FCS) \cite{FuzzyCommitment}, which are based on a one-time padding step in combination with an error correcting code. Both constructions require a storage rate of $1$ bit/symbol due to the one-time padding step.
A Slepian-Wolf (SW) \cite{SW} coding method, which corresponds to syndrome coding for binary sequences, is proposed in \cite{IgnaPolar} to reduce the storage rate so that it is equal to the privacy-leakage rate.
It is shown in \cite{OurWZTrans} that these methods do not achieve the key-leakage-storage boundaries of the GS and CS models.

Wyner-Ziv (WZ) \cite{WZ} coding constructions that bin the observed sequences are shown in \cite{OurWZTrans} to be optimal deterministic code constructions for key agreement with PUFs.
Nested random linear codes are shown to asymptotically achieve boundary points of the key-leakage-storage region.
A second WZ-coding construction uses a nested version of polar codes (PCs) \cite{Arikan}, which are designed in \cite{OurWZTrans} for practical SRAM PUF parameters to illustrate that rate tuples that cannot be achieved by using previous code constructions can be achieved by nested PCs.  

A closely related problem to the key agreement problem is Wyner's wiretap channel (WTC) \cite{WynerWTC}. The main aim in the WTC problem is to hide a transmitted message from the eavesdropper that observes a channel output correlated with the observation of a legitimate receiver. There are various code constructions for the WTC that achieve the secrecy capacity, e.g., in \cite{WTCpolarVardy,KliewerWTC,OzanWTC,YingbingWTC}, and some of these constructions use nested PCs, e.g., \cite{KliewerWTC,YingbingWTC}. Similarly, nested PCs are shown in \cite{RemiStrongCoordination} to achieve the strong coordination capacity boundaries, defined and characterized in \cite{strongcoordinationPolar}. 

We design codes for key agreement with PUFs by constructing nested convolutional codes. Due to the broad use of nested codes in, e.g., WTC and strong coordination problems, the proposed nested convolutional code constructions can be useful also for these problems. A summary of the main contributions is as follows.

\begin{itemize}	
	\item We propose a method to obtain nested tailbiting convolutional codes (TBCCs) that are used as a WZ-coding construction, which is a binning method used in various achievability schemes and can be useful for various practical code constructions.  
	\item We develop a design procedure for the proposed nested convolutional code construction adapted to the problem of key agreement with biometric or physical identifiers. This is an extension of the asymptotically optimal nested code constructions with random linear codes and PCs proposed in \cite{OurWZTrans}. We consider binary symmetric sources and binary symmetric channels (BSCs). Physical identifiers such as RO PUFs with transform coding \cite{OurEntropy} and SRAM PUFs \cite{maes2009soft} are modeled by these sources and channels.
	\item We design and simulate nested TBCCs for practical source and channel parameters obtained from the best PUF design in the literature. The target block-error probability is $P_B=10^{-6}$ and the target secret-key size is 128 bits. We illustrate that one variant of nested codes achieves the largest key vs. storage rate ratio but it has high decoding complexity. Another variant of nested codes with lower decoding complexity achieves a rate ratio that is slightly greater than the rate ratio achieved by a nested PC. We also illustrate the gaps to the finite-length bounds.
\end{itemize}

This paper is organized as follows. In Section~\ref{sec:problem_settingandcode}, we describe the GS and CS models, and give their rate regions that are also evaluated for binary symmetric sequences. We summarise in Section~\ref{sec:nestedconv} our new nested code construction that uses convolutional codes. In Section~\ref{sec:nestedconvforPUFs}, we propose a design procedure for the new nested TBCCs adapted to the key agreement with PUFs problem. Section~\ref{sec:decCompEst} compares the estimated decoding complexity of TBCCs and PCs. Section~\ref{sec:results} illustrates the significant gains from nested convolutional codes designed for practical PUF parameters as compared to previously-proposed nested PCs and other channel codes in terms of the key vs. storage rate ratio.

\section{Preliminaries}
\subsection{Notation}
Let $\Fq$ denote the finite field of order $2$ and let $\Fq^{a \times b}$ denote the set of all $a \times b$ matrices over $\Fq$. Rows and columns of $a \times b$ matrices are indexed by $1,\ldots,a$ and $1,\ldots,b$, and $h_{i,j}$ is the element in the $i$-th row and $j$-th column of a matrix $\mathbf{H}$. $\Fq^{a}$ denotes the set of all row vectors of length $a$ over $\Fq$. With $\bm{0}_{a \times b}$ we denote the all-zero matrix of size $a \times b$. A linear block code over $\Fq$ of length $N$ and dimension $K$ is a $K$-dimensional subspace of $\Fq^{N}$ and denoted by $(N,K)$. A variable with superscript denotes a string of variables, e.g., $\displaystyle X^n\!=\!X_1\ldots X_i\ldots X_n$, and a subscript denotes the position of a variable in a string. A random variable $\displaystyle X$ has probability distribution $\displaystyle P_X$. Calligraphic letters such as $\displaystyle \mathcal{X}$ denote sets, and set sizes are written as $\displaystyle |\mathcal{X}|$. $\Enc(\cdot)$ is an encoder mapping and $\Dec(\cdot)$ is a decoder mapping. $H_b(x)=-x\log x- (1-x)\log (1-x)$ is the binary entropy function, where we take logarithms to the base $2$. The $*$-operator is defined as $\displaystyle p*x = p(1-x)+(1-p)x$. A BSC with crossover probability $p$ is denoted by BSC($p$). $X^n\sim\text{Bern}^n(\alpha)$ is an independent and identically distributed (i.i.d.) binary sequence of random variables with $\Pr[X_i=1]=\alpha$ for $i=1,2,\ldots,n$. ${\mathbf{H}}^T$ represents the transpose of the matrix $\mathbf{H}$.
Drawing an element $e$ from a set $\mathcal{E}$ uniformly at random is denoted by
\begin{equation}
 e \assignRand \mathcal{E}.   
\end{equation}

\subsection{Convolutional Codes}\label{sec:tbcc_prelim}

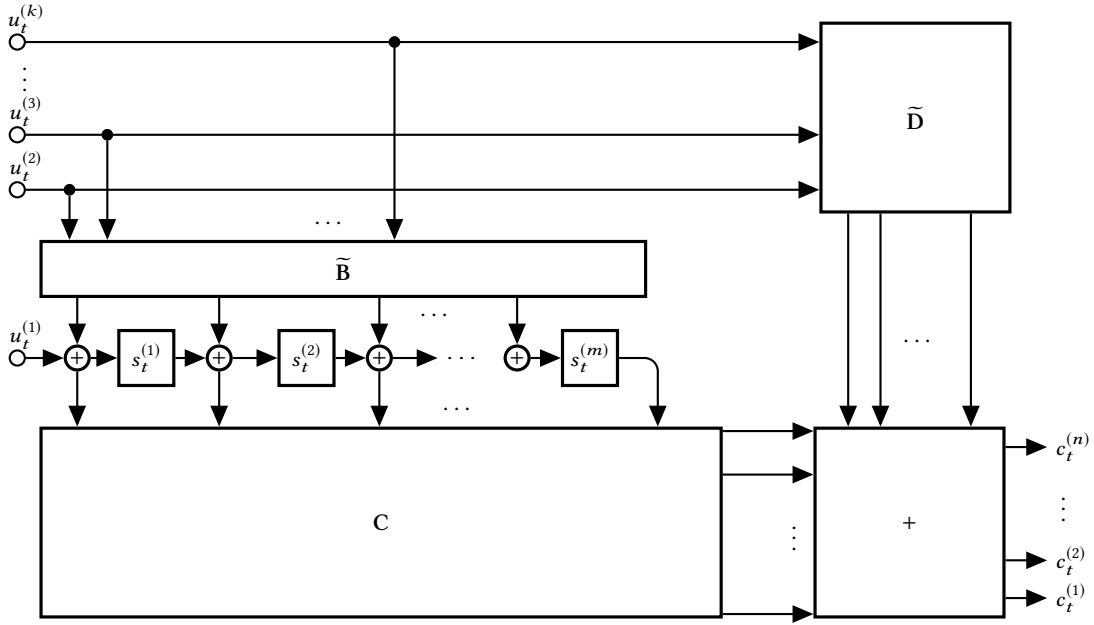
\begin{figure*}
	\centering
	\resizebox{0.83\linewidth}{!}{
    	\begin{tikzpicture}[
    	node distance=0.75cm,
    	>=triangle 45,
    	]
    	\tikzset{
    		delayCell/.style={rectangle,draw=black, very thick, inner sep=0.25em, minimum size=2.3em, text centered},
    		genMatrix/.style={rectangle,draw=black, very thick, inner sep=0.0em, minimum size=3.0em, text centered},
    		sumNode/.style={circle,draw=black, very thick, inner sep=0.0em, minimum size=1.0em, text centered, node distance=0.25cm},
    		multNode/.style={circle,draw=black, very thick, inner sep=0.0em, minimum size=1.0em, text centered, node distance=0.25cm},	
    		conArrow/.style={->, thick,rounded corners=5pt},
    		conLine/.style={-, thick,rounded corners=5pt},	
    		con/.style={thick,rounded corners=5pt},
    		inputNode/.style={circle, draw=black, thick, inner sep=0.0em, minimum size=2mm},
    		conNode/.style={circle,draw=black, fill=black, minimum size=4pt, inner sep=0.0em},
    	}
    	
    	\node[inputNode] (input0) at (0,0){};
    	
    	\node[inputNode, above=2cm of input0] (input1) {};
    	\node[inputNode, above=0.5cm of input1] (input2) {};
    	\node[inputNode, above=1cm of input2] (input3) {};
    	
    	\node[conNode, right=0.50cm of input1] (con2) {};
    	\node[conNode, right=1.0cm of input2] (con5) {};
    	\node[conNode, right=4.8cm of input3] (con8) {};
    	
    	\node[sumNode, right=0.50cm of input0] (fbSum) {$+$};
    	
    	\node[delayCell, right=0.35cm of fbSum] (dc00) {\small $s_t^{(1)}$};
    	\node[delayCell, right=1.35cm of dc00] (dc01) {\small $s_t^{(2)}$};
    	\node[right=1.35cm of dc01] (dcxx) {\small $\ldots$};
    	\node[delayCell, right=1.0cm of dcxx] (dc02) {\small $s_t^{(m)}$};
    	
    	\node[delayCell, draw=none, right=1cm of dc02] (dc03) {};

    	\node[sumNode, right=0.4cm of dc00] (con3) {$+$};
    	\node[sumNode, right=0.40cm of dc01] (con6) {$+$};
    	\node[right=0.40cm of dc02] (con9) {};
    	\node[sumNode, left=0.40cm of dc02] (con12) {$+$};
    	
    	\node[delayCell, above=1cm of fbSum, minimum width=8cm, anchor=west, xshift=-0.5cm] (Bmatrix) {$\mathbf{\widetilde{B}}$};
    	\node[delayCell, below=2cm of fbSum, minimum width=9cm, minimum height=2.5cm, anchor=west, xshift=-0.5cm] (Cmatrix) {$\mathbf{C}$};
    	\node[delayCell, right=11cm of input3, minimum size=2.5cm, yshift=-1cm, anchor=west, xshift=-0.5cm] (Dmatrix) {$\mathbf{\widetilde{D}}$};
    	\node[delayCell, right=1.2cm of Cmatrix, minimum size=2.5cm, anchor=west] (Plusmatrix) {$+$};
    	
    	\node[right=0.55cm of Plusmatrix, yshift=1cm] (out0) {\small $c_t^{(n)}$};
    	\node[right=0.55cm of Plusmatrix, yshift=-0.5cm] (out1) {\small $c_t^{(2)}$};
    	\node[right=0.55cm of Plusmatrix, yshift=-1cm] (out2) {\small $c_t^{(1)}$};
    	
    	\node[above=0.1cm of out1, xshift=-0.1cm] {$\vdots$};
    	\node[left=0.1cm of Plusmatrix, yshift=-0.1cm] {$\vdots$};
    	\node[above=1cm of Plusmatrix, xshift=0.1cm] {$\ldots$};
    	\node[below=0.0cm of input3, xshift=0.1cm, yshift=0.1cm] {$\vdots$};
    	
    	\node[below=0.1cm of Bmatrix, xshift=1.2cm] {$\ldots$};
    	\node[above=0.1cm of Bmatrix, xshift=-0.2cm] {$\ldots$};
    	\node[above=0.1cm of Cmatrix, xshift=1cm] {$\ldots$};
    	
    	\draw[conArrow] (Plusmatrix.east |- out0) -- (out0);
    	\draw[conArrow] (Plusmatrix.east |- out1) -- (out1);
    	\draw[conArrow] (Plusmatrix.east |- out2) -- (out2);
    	
    	\draw[conArrow] (input0) -- (fbSum) node[at start, above]{\small $u_t^{(1)}$};
    	\draw[conArrow] (fbSum) -- (dc00);
    	\draw[conArrow] (dc00) -- (con3);
    	\draw[conArrow] (con3) -- (dc01);
    	\draw[conArrow] (dc01) -- (con6);
    	\draw[conArrow] (con6) -- (dcxx);
    	\draw[conArrow] (con12) -- (dc02);
    	
    	\draw[conArrow] (Bmatrix.south -| fbSum.north) -- (fbSum);
    	\draw[conArrow] (Bmatrix.south -| con3.north) -- (con3);
    	\draw[conArrow] (Bmatrix.south -| con6.north) -- (con6);
    	\draw[conArrow] (Bmatrix.south -| con12.north) -- (con12);
    	
    	\draw[conArrow] (fbSum) -- (Cmatrix.north -| fbSum.south);
    	\draw[conArrow] (con3) -- (Cmatrix.north -| con3.south);
    	\draw[conArrow] (con6) -- (Cmatrix.north -| con6.south);
    	\draw[conArrow] (dc02) -| (Cmatrix.north -| con9.south);
    	
    	\draw[conArrow] (con2) -- (Bmatrix.north -| con2.south);
    	\draw[conArrow] (con5) -- (Bmatrix.north -| con5.south);
    	\draw[conArrow] (con8) -- (Bmatrix.north -| con8.south);
    	
    	\draw[conArrow] (input1) -- (input1.west -| Dmatrix.west) node[at start, above]{\small $u_t^{(2)}$};
    	\draw[conArrow] (input2) -- (input2.west -| Dmatrix.west) node[at start, above]{\small $u_t^{(3)}$};
    	\draw[conArrow] (input3) -- (input3.west -| Dmatrix.west) node[at start, above]{\small $u_t^{(k)}$};
    	
    	\draw[conArrow] (Cmatrix.15) -- (Plusmatrix.west |- Cmatrix.15);
    	\draw[conArrow] (Cmatrix.8) -- (Plusmatrix.west |- Cmatrix.8);
    	\draw[conArrow] (Cmatrix.-15) -- (Plusmatrix.west |- Cmatrix.-15);
    	
    	\draw[conArrow] (Dmatrix.-125) -- (Plusmatrix.north -| Dmatrix.-125);
    	\draw[conArrow] (Dmatrix.-110) -- (Plusmatrix.north -| Dmatrix.-110);
    	\draw[conArrow] (Dmatrix.-60) -- (Plusmatrix.north -| Dmatrix.-60);
    	
    	\end{tikzpicture}
    }
	\caption{Encoder circuit of convolutional codes described in Section~\ref{sec:tbcc_prelim}.}\label{fig:encoderFig}
\end{figure*}

Denote the parameters of a block code generated by a binary convolutional encoder as $(N,K)$, where $N$ is the blocklength and $K$ is the code dimension (in bits). At each time step, the convolutional encoder receives $k$ input bits and generates $n$ output bits. The number of clock cycles needed to encode $K$ bits is $\ell=\frac{K}{k}$. We consider convolutional encoders with a single shift register only. The shift register consists of $m$ delay cells, where $m$ is also called the memory of the encoder. The bit value stored in the $i$-th delay cell at time step $t$ is denoted by $s_t^{(i)} \in \Fq$ for $i=1,\ldots,m$. For a given binary input vector $\mathbf{u}_t = \left(u_t^{(1)},u_t^{(2)},\ldots,u_t^{(k)}\right)$ of length $k$ at time step $t$, the encoder outputs a binary vector $\mathbf{c}_t = \left(c_t^{(1)},c_t^{(2)},\ldots,c_t^{(n)}\right)$ of length $n$. The encoder can be described by the state-space representation of the encoder circuit such that the output $\mathbf{c}_t$ is 
\begin{equation}
  \mathbf{c}_t =  \mathbf{s}_t \cdot \mathbf{C}^{T} +  \mathbf{u}_t \cdot \mathbf{D}^{T}
\end{equation}
where $\mathbf{s}_t = \left(s_t^{(1)},s_t^{(2)},\ldots,s_t^{(m)}\right)$ is the vector describing the content of the shift register, $\mathbf{C} \in \Fq^{n\times m}$ is the observation matrix, and $\mathbf{D} \in \Fq^{n \times k}$ is the transition matrix.
The content of the shift register for the next clock cycle at time step $t+1$ is then
\begin{equation}
  \mathbf{s}_{t+1} =  \mathbf{s}_t \cdot \mathbf{A}^{T} +  \mathbf{u}_t \cdot \mathbf{B}^{T}
\end{equation}
where $\mathbf{A} \in \Fq^{m\times m}$ is the system matrix and $\mathbf{B} \in \Fq^{m \times k}$ is the control matrix. For the case of a single shift register we have that the system matrix is given by
\begin{equation}
  \mathbf{A} = \begin{bmatrix}
  \bm{0}_{1 \times (m-1)} & 0  \\
  \mathbf{I}_{(m-1)\times(m-1)} & \bm{0}_{(m-1) \times 1} \\
  \end{bmatrix}
\end{equation}
where $\mathbf{I}_{(m-1)\times(m-1)} \in \Fq^{(m-1)\times(m-1)}$ is the identity matrix. For simplicity, first entry of the input tuple $u_t^{(1)}$ is always an input to the shift register and thus we can write $\mathbf{B} = (\mathbf{e}_1^{T} | \widetilde{\mathbf{B}})$ and $\mathbf{D} = (\bm{0}_{n \times 1} | \widetilde{\mathbf{D}})$, where $\mathbf{e}_1$ is the unit row vector having a $1$ in the first position and $0$ everywhere else, $\widetilde{\mathbf{B}} \in \Fq^{m\times(k-1)}$, and $\widetilde{\mathbf{D}} \in \Fq^{n\times(k-1)}$. The corresponding encoder circuit is shown in Figure~\ref{fig:encoderFig}. Elements of a vector entering a square box, which represents one of the aforementioned matrices, depicts a vector-matrix multiplication, and the box with the addition symbol depicts an elementwise vector-vector addition. Therefore, the encoder of the convolutional code can be described by the three matrices $\widetilde{\mathbf{B}}$, $\mathbf{C}$, and $\widetilde{\mathbf{D}}$. We denote such an encoder by $\ccenc{\widetilde{\mathbf{B}}}{\mathbf{C}}{\widetilde{\mathbf{D}}}$.

Using the tailbiting method from \cite[Chapter~4.8]{JZ15}, we avoid having a rate loss, unlike the zero-tail termination method. We have $N = \ell n$ and the resulting code rate is $R=\frac{k}{n}$. A \textit{tailbiting convolutional code (TBCC)} can be represented by a tailbiting trellis using $\ell$ sections and $2^m$ states per section. The codewords correspond to all possible paths in the trellis, where starting and ending states coincide. TBCCs can be decoded by using the wrap around Viterbi algorithm (WAVA) \cite{wava03}. This decoder is suboptimal but performs close to the performance of the maximum likelihood decoder.

Let $A_d$ be the number of codewords of Hamming weight $d$ for $d=0,1,\ldots,N$, which characterizes the distance spectrum of a TBCC. The weight enumerator polynomial $A(X)$ is then defined as
\begin{equation}\label{eq:wef}
    A(X) \; \eqdef \; \sum_{d=0}^{N} A_d X^{d}.
\end{equation}
To compute the weight enumerator and to determine the distance spectrum we use the approach described in \cite{WV96}. Consider the state transition matrix $\mathbf{T}(X)$ of size $2^m \times 2^m$, where every entry $t_{i,j}(X)$ is either $X^d$, where $d$ is the Hamming weight of the output produced by the encoder when going from the state labeled with $i$ to the state labeled with $j$, or $0$ if there is no possible transition between the aforementioned states. Therefore, we have
\begin{equation}\label{eq:wef_comp}
  A(X) = \text{Tr}\left({\mathbf{T}^{\ell}(X)}\right)
\end{equation}
where $\mathbf{T}^{\ell}(X)$ denotes multiplication of the matrix $\mathbf{T}(X)$ with itself $\ell$ times and $\text{Tr}(\cdot)$ denotes the trace.

\section{Problem Formulation}\label{sec:problem_settingandcode}
\begin{figure}
	\centering
	\resizebox{0.99\linewidth}{!}{
        \begin{tikzpicture}
        \node (so) at (-1.5,-2.2) [draw,rounded corners = 5pt, minimum width=1.0cm,minimum height=0.8cm, align=left] {$P_X(\cdot)$};
        \node (a) at (0,0) [draw,rounded corners = 6pt, minimum width=3.2cm,minimum height=1.2cm, align=left] {$
        	(S,W) \overset{(a)}{=} \Enc\left(X^N\right)$\\ $W'\overset{(b)}{=}\Enc\left(X^N,S'\right)$};
        \node (c) at (3,-2.2) [draw,rounded corners = 5pt, minimum width=1.6cm,minimum height=0.8cm, align=left] {$P_{Y|X}(\cdot)$};
        \node (b) at (6,0) [draw,rounded corners = 6pt, minimum width=3.2cm,minimum height=1.2cm, align=left] {$\widehat{S} \overset{(a)}{=} \Dec\left(Y^N,W\right)$\\$\widehat{S}' \overset{(b)}{=} \Dec\left(Y^N,W'\right)$};
        \draw[decoration={markings,mark=at position 1 with {\arrow[scale=1.5]{latex}}},
        postaction={decorate}, thick, shorten >=1.4pt] (a.east) -- (b.west) node [midway, above] {$(a) W$} node [midway, below] {$(b) W'$};;
        \node (a1) [below of = a, node distance = 2.2cm] {$X^N$};
        \node (b1) [below of = b, node distance = 2.2cm] {$Y^N$};
        \node (k9) [below of = a1, node distance = 0.6cm] {Enrollment};
        \node (k19) [below of = b1, node distance = 0.6cm] {Reconstruction};
        \draw[decoration={markings,mark=at position 1 with {\arrow[scale=1.5]{latex}}},
        postaction={decorate}, thick, shorten >=1.4pt] (so.east) -- (a1.west);
        \draw[decoration={markings,mark=at position 1 with {\arrow[scale=1.5]{latex}}},
        postaction={decorate}, thick, shorten >=1.4pt] (a1.north) -- (a.south);
        \draw[decoration={markings,mark=at position 1 with {\arrow[scale=1.5]{latex}}},
        postaction={decorate}, thick, shorten >=1.4pt] (a1.east) -- (c.west);
        \draw[decoration={markings,mark=at position 1 with {\arrow[scale=1.5]{latex}}},
        postaction={decorate}, thick, shorten >=1.4pt] (c.east) -- (b1.west);
        \draw[decoration={markings,mark=at position 1 with {\arrow[scale=1.5]{latex}}},
        postaction={decorate}, thick, shorten >=1.4pt] (b1.north) -- (b.south);
        \node (a2) [above of = a, node distance = 2.2cm] {$S\quad\,\, S'$};
        \node (b2) [above of = b, node distance = 2.2cm] {$\widehat{S}\quad\,\,\widehat{S}'$};
        \draw[decoration={markings,mark=at position 1 with {\arrow[scale=1.5]{latex}}},
        postaction={decorate}, thick, shorten >=1.4pt] ($(b.north)-(0.3,0)$) -- ($(b2.south)-(0.3,0)$) node [midway, left] {$(a)$};
        \draw[decoration={markings,mark=at position 1 with {\arrow[scale=1.5]{latex}}},
        postaction={decorate}, thick, shorten >=1.4pt] ($(b.north)+(0.3,0)$)-- ($(b2.south)+(0.3,0)$) node [midway, right] {$(b)$};
        \draw[decoration={markings,mark=at position 1 with {\arrow[scale=1.5]{latex}}},
        postaction={decorate}, thick, shorten >=1.4pt] ($(a.north)-(0.3,0)$)-- ($(a2.south)-(0.3,0)$) node [midway, left] {$(a)$};
        \draw[decoration={markings,mark=at position 1 with {\arrow[scale=1.5]{latex}}},
        postaction={decorate}, thick, shorten >=1.4pt]  ($(a2.south)+(0.3,0)$)-- ($(a.north)+(0.3,0)$) node [midway, right] {$(b)$};
        \end{tikzpicture}
	}
	\caption{The $(a)$ GS and $(b)$ CS models.}\label{fig:problemsetup}
\end{figure}
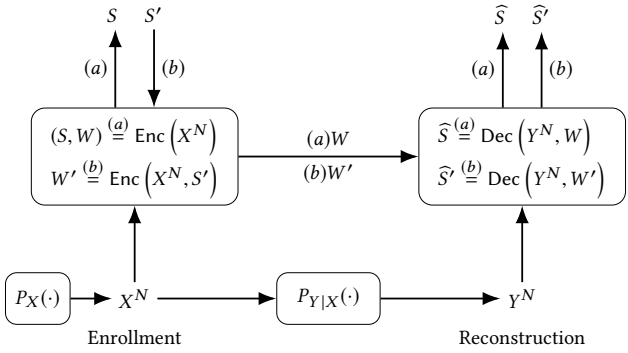

Consider the GS model in Figure~\ref{fig:problemsetup}$(a)$, where a biometric or physical source output is used to generate a secret key. The source $\mathcal{X}$, noisy measurement $\mathcal{Y}$, secret key $\mathcal{S}$, and storage $\mathcal{W}$ alphabets are finite sets. During enrollment, the encoder observes the i.i.d. identifier output $X^N$, generated according to some $P_X$, and computes a secret key $S\in\mathcal{S}$ and public helper data $W\in\mathcal{W}$ as $\displaystyle (S,W)\,{=}\,{\Enc}(X^N)$. During reconstruction, the decoder observes a noisy source measurement $Y^N$ of the source output $X^N$ through a memoryless measurement channel $P_{Y|X}$ in addition to the helper data $W$. The decoder estimates the secret key as $\displaystyle \widehat{S}\,{=}\,{\Dec}(Y^N\!,W)$. Furthermore, Figure~\ref{fig:problemsetup}$(b)$ shows the CS model, where a secret key $S'\in\mathcal{S}$ is embedded into the helper data as $W' = \Enc(X^N,S')$. The decoder for the CS model estimates the secret key as $\widehat{S}'=\Dec(Y^N,W')$.

\begin{definition}\label{def:achievabilityGSCS}
	 A key-leakage-storage tuple $(R_s,R_\ell,R_w)$ is \emph{achievable} for the GS and CS models if, given any $\epsilon>0$, there is some $N\!\geq\!1$, an encoder, and a decoder such that $R_s=\frac{\log|\mathcal{S}|}{N}$ and
	\begin{align}
	&\pB\;\eqdef\;\Pr[\widehat{S} \neq S] \leq \epsilon&&\quad (\text{reliability})\label{eq:reliability_constraint}\\
	&\frac{1}{N}I(S;W) \leq \epsilon&&\quad(\text{secrecy})\label{eq:secrecyleakage_constraint}\\
	&\frac{1}{N}H(S)\geq R_s-\epsilon&&\quad(\text{key uniformity}) \label{eq:uniformity_constraint}\\
	&\frac{1}{N}\log\big|\mathcal{W}\big| \leq R_w+\epsilon&&\quad(\text{storage})\label{eq:storage_constraint}\\
	&\frac{1}{N}I(X^N;W) \leq R_\ell+\epsilon&&\quad(\text{privacy})\label{eq:leakage_constraint}
	\end{align}
	where for the CS model, $S$ and $W$ in the constraints should be replaced by, respectively, $S'$ and $W'$.
	 
    The \emph{key-leakage-storage} regions $\mathcal{R}_{\text{gs}}$ and $\mathcal{R}_{\text{cs}}$ for the GS and CS models, respectively, are the closures of the sets of achievable tuples for the corresponding models.\hfill $\lozenge$
\end{definition}

\begin{theorem}[\hspace{1sp}\cite{IgnaTrans}]\label{theo:secrecyregions}
The key-leakage-storage region $\mathcal{R}_{\text{gs}}$ for the GS model is the union of the bounds 
\begin{align}
&0 \leq R_s\leq I(U;Y)\label{eq:Rsregion}\\
&R_\ell\geq I(U;X)-I(U;Y)\label{eq:Rlregion}\\
&R_w\geq I(U;X)-I(U;Y)\label{eq:RWregionGS}
\end{align}
over all $P_{U|X}$ such that $U-X-Y$ form a Markov chain. Similarly, the key-leakage-storage region $\mathcal{R}_{\text{cs}}$ for the CS model is the union of the bounds in (\ref{eq:Rsregion}), (\ref{eq:Rlregion}), and
\begin{align}
	&R_w\geq I(U;X)\label{eq:RWregionCS}.
\end{align}
These regions are convex sets. The alphabet $\mathcal{U}$ of the auxiliary random variable $U$ can be limited to have size $\displaystyle |\mathcal{U}|\!\leq\!|\mathcal{X}|+1$. Deterministic encoders and decoders suffice to achieve these regions.
\end{theorem}

Suppose the transform-coding algorithms proposed in \cite{OurEntropy} are applied to RO PUFs or any PUF circuits with continuous-valued outputs to obtain $X^N$ that is almost i.i.d. according to a uniform Bernoulli random variable, i.e., $X^N\sim \text{Bern}^N(\frac{1}{2})$, and the channel $P_{Y|X}$ is a $\text{BSC}(p_A)$ for $p_A\in[0, 0.5]$. The key-leakage-storage region $\mathcal{R}_{\text{gs,bin}}$ of the GS model for this case is the union of the bounds 
\begin{align}
&0\leq R_s\leq 1- H_b(q*p_A)\nonumber\\
&R_\ell\geq H_b(q*p_A)- H_b(q)\nonumber\\
&R_w\geq H_b(q*p_A)- H_b(q)\label{eq:BSCRegionGS}
\end{align}
over all $q\in[0,0.5]$ \cite{IgnaTrans}, which follows by using an auxiliary random variable $U$ such that $P_{X|U}\sim\text{BSC}(q)$ due to Mrs. Gerber's lemma \cite{WZ}. The rate tuples on the boundary of the region $\mathcal{R}_{\text{gs,bin}}$ are uniquely defined by the ratio $\displaystyle\frac{R_s}{R_w}$. We therefore use this ratio as the metric to compare our nested TBCCs with previously-proposed nested PCs and channel codes. A larger key vs. storage rate ratio suggests that the code construction is closer to an achievable point that is on the boundary of the region $\mathcal{R}_{\text{gs,bin}}$, which is an optimal tuple. We next focus on the GS model for code constructions. All results can be extended to the CS model by using an additional one-time padding step \cite{benimdissertation}.


\section{Nested Convolutional Code Construction}\label{sec:nestedconv}
In this section, we sketch the main steps to obtain a nested construction for convolutional codes. Furthermore, we give two explicit algorithms to find good code constructions. The first algorithm addresses the search of a good error correcting code $(N,\fecDim)$, denoted by $\fecCode$, and the second algorithm finds a $(N,\quantDim)$ code $\quantCode$ used as a vector quantizer such that $\fecCode$ is a subcode of $\quantCode$, i.e., $\fecCode \subseteq \quantCode$.

\subsection{Nested Convolutional Codes}
Using the encoder circuit depicted in Figure~\ref{fig:encoderFig}, we construct two codes $\quantCode$ and $\fecCode$ such that $\fecCode \subseteq \quantCode$.
Let $\quantCode$ be the $(N,\quantDim)$ TBCC with memory $m$ and $\quantDim=\ell\quantk$ generated by using the encoder defined by the matrices $\ccenc{\widetilde{\mathbf{B}}}{\widetilde{\mathbf{C}}}{\widetilde{\mathbf{D}}}$. Recall that $\mathbf{B} = (\mathbf{e}_1^{T} | \widetilde{\mathbf{B}})$ with $\widetilde{\mathbf{B}} \in \Fq^{m \times (\quantk-1)}$ and $\mathbf{D} = (\bm{0}^T | \widetilde{\mathbf{D}})$ with $\widetilde{\mathbf{D}} \in \Fq^{n \times (\quantk-1)}$. By removing the $i$-th column of $\widetilde{\mathbf{B}}$ and $\widetilde{\mathbf{D}}$ simultaneously, one obtains a new encoder that generates a code of rate $\displaystyle\frac{\quantk-1}{n}$, which is a subcode of the original code. This is true, since the new code corresponds to all codewords by encoding the original code but restricting to all inputs where $u_t^{(i)} = 0$.
By ``freezing'' further input bits we can therefore obtain a subcode of rates
\begin{equation}
 \fecR = \dfrac{1}{n}, \dfrac{2}{n}, \ldots,\dfrac{\quantk-1}{n}. \label{eq:listofintegernumrates}
\end{equation}
To obtain codes with rates of better granularity between $\dfrac{1}{n}$ and $\dfrac{\quantk-1}{n}$ that are not in \eqref{eq:listofintegernumrates}, we can freeze input bits in a time-variant manner. That is, by using the encoder $\ell$ times, we can freeze a different amount of input bits in different clock cycles. This allows to obtain codes of rates
\begin{equation}
\fecR = \dfrac{\ell}{N}, \dfrac{\ell+1}{N}, \ldots,\dfrac{\quantDim}{N}.
\end{equation}
Denote the parameters of the subcode, obtained by freezing input bits accordingly, as $(N,\fecDim)$. Note that by freezing input bits in a time-variant manner, $\fecDim$ is not necessarily a multiple of $\ell$. Furthermore, the procedure can be applied also to add columns to $\widetilde{\mathbf{B}}$ and $\widetilde{\mathbf{D}}$ to generate a supercode. The design procedure of the nested convolutional code construction is split into two steps:
\begin{enumerate}
    \item Search for a good error correcting code $\fecCode$ of rate $\fecR=\dfrac{1}{n}=\dfrac{\fecDim}{N}$ at given target block error probability $\pB$ by finding an appropriate matrix $\mathbf{C}$.
    \item Expand the low rate code by finding appropriate matrices $\widetilde{\mathbf{B}}$ and $\widetilde{\mathbf{D}}$ to obtain a good code of rate $\quantR=\dfrac{\quantk}{n}=\dfrac{\quantDim}{N}$ that achieves a low average distortion $q$.
\end{enumerate}
Note that for the first step we restrict to codes of rate $\fecR=\dfrac{1}{n}$ and hence the matrices $\widetilde{\mathbf{B}}$ and $\widetilde{\mathbf{D}}$ are vanishing. The first step can also be performed for codes of any rate $\fecR>\dfrac{1}{n}$, but then also the appropriate matrices $\widetilde{\mathbf{B}}$ and $\widetilde{\mathbf{D}}$ have to be found accordingly.

\begin{algorithm}[t]
	\caption{Search for $(N,\fecDim)$ TBCC $\fecCode$, $\fecR=\dfrac{1}{n}$}\label{alg:designFecCode}
	\SetKwInOut{Input}{Input}\SetKwInOut{Output}{Output}
	\Input{$n$, $m$, $\fecDim$, $\pB$, $W_\text{max}$ (maximum number of iterations)}
	\Output{$\mathbf{C} \in \Fq^{n \times m}$}
    \BlankLine
    \textbf{Initialize:} \\
    $\pc \leftarrow 0$\\
    $\mathbf{C} \leftarrow \bm{0}$\\
	\BlankLine
	\For{$w\leftarrow 1$ \KwTo $W_\text{max}$}{
    	$\mathbf{C}' \assignRand \Fq^{n \times m} $ \\
    	Compute $A_d$ for the $(N,\fecDim)$ TBCC generated by $\ccenc{\mathbf{0}}{\mathbf{C}'}{\mathbf{0}}$ for $d=0,\ldots,N$ using~\eqref{eq:wef} and~\eqref{eq:wef_comp}\\
    	Find $\pc'$ such that: $\UB(A_d,\pc') = \pB$ \\
    	\If{$\pc' \geq \pc$}{
    	    $\pc \leftarrow \pc'$ \\
    	    $\mathbf{C} \leftarrow \mathbf{C}'$
    	}
	}
	\Return{$\mathbf{C}$}
\end{algorithm}
\subsection{Design of a Convolutional Code for Error Correction}
For fixed parameters $n$, $m$, and $\fecDim$, we try to find a matrix $\mathbf{C}$ such that the resulting $(N,\fecDim)$ TBCC $\fecCode$ at a given target block error probability $\pB$ can be operated on a noisy BSC with large crossover probability $\pc$. To evaluate $\pB$ we use the union bound, see, e.g., \cite{Polty94}, and the distance spectrum of the code. This gives an upper bound on $\pB$ under maximum likelihood decoding. The bound is given by 
\begin{equation}\label{eq:ubbound}
  \pB \leq \UB(A_d,\pc) \; \eqdef \sum_{d=\dmin}^{N} A_d \sum_{i=\lceil d/2 \rceil}^{d} \binom{d}{i}  \pc^i (1-\pc)^{d-i}
\end{equation}
where $\dmin$ is the minimum distance of the code.

The design of the code $\fecCode$ is performed by a purely random search of the matrix $\mathbf{C}$ as described in Algorithm~\ref{alg:designFecCode}. This algorithm searches the best TBCC of rate $\fecR=\dfrac{1}{n}$ by randomly generating different matrices $\mathbf{C}$. The matrix $\mathbf{C}$ of the code that yields the largest $\pc$ at a given target block error probability $\pB$ is returned as the output of Algorithm~\ref{alg:designFecCode}.

\begin{algorithm}[t]
	\caption{Search for $(N,\quantDim)$ TBCC $\quantCode$, $\quantR=\dfrac{\quantk}{n}$}\label{alg:designVQCode}
	\SetKwInOut{Input}{Input}\SetKwInOut{Output}{Output}
	\Input{$m$, $\quantk$, $\feck$, $W_\text{max}$, $\mathbf{C}$, $\widetilde{\mathbf{B}}_\fec \in \Fq^{m \times (\feck-1)}$, $\widetilde{\mathbf{D}}_\fec \in \Fq^{n \times (\feck-1)}$}
	\Output{$\widetilde{\mathbf{B}}_\vq \in \Fq^{m \times (\quantk-1)}$, $\widetilde{\mathbf{D}}_\vq \in \Fq^{n \times (\quantk-1)}$}
    \BlankLine
    \textbf{Initialize:} \\
    $\widetilde{\mathbf{B}}_\vq \leftarrow (\widetilde{\mathbf{B}}_\fec | \bm{0})$ \\
    $\widetilde{\mathbf{D}}_\vq \leftarrow (\widetilde{\mathbf{D}}_\fec | \bm{0})$ \\
    $d \leftarrow 0$ \\
    $A \leftarrow 0$ \\
	\BlankLine
	\For{$w\leftarrow 1$ \KwTo $W_\text{max}$}{
	    $\mathbf{B}' \assignRand \Fq^{m \times (\quantk-\feck)}$ \\
	    $\mathbf{D}' \assignRand \Fq^{n \times (\quantk-\feck)}$ \\
	    $\widetilde{\mathbf{B}}'_\vq \leftarrow (\widetilde{\mathbf{B}}_\fec | \mathbf{B}')$ \\
        $\widetilde{\mathbf{D}}'_\vq \leftarrow (\widetilde{\mathbf{D}}_\fec | \mathbf{D}')$ \\
        Compute $\dfree$ and $\Afree$ for $\ccenc{\widetilde{\mathbf{B}}_\vq}{\mathbf{C}}{\widetilde{\mathbf{D}}_\vq}$\\
    	\If{$\dfree > d$ or ($\dfree = d$ and $\Afree < A$)}{
            $d \leftarrow \dfree$ \\
            $A \leftarrow \Afree$ \\
            $\widetilde{\mathbf{B}}_\vq \leftarrow \widetilde{\mathbf{B}}'_\vq$ \\
            $\widetilde{\mathbf{D}}_\vq \leftarrow \widetilde{\mathbf{D}}'_\vq$ \\
    	}
	}
	\Return{$\widetilde{\mathbf{B}}_\vq$, $\widetilde{\mathbf{D}}_\vq$}
\end{algorithm}

\subsection{Design of a Convolutional Code for Vector Quantization}

In this section, an algorithm to obtain a high rate code from an existing low rate convolutional encoder is explained.
The algorithm is presented in Algorithm~\ref{alg:designVQCode}.
The inputs are the system matrix, the observation matrix, and the transition matrix of the low rate code with rate
\begin{align}
 \fecR=\frac{\feck}{n}.   
\end{align}
By randomly adding $\quantk-\feck$ columns to both, the system and the transition matrix of a code of high rate
\begin{align}
    \quantR=\frac{\quantk}{n}    
\end{align}
is constructed. The algorithm performs a random search and returns the best configuration. As selection metrics, the free distance and its multiplicity are chosen. The free distance $\dfree$ of a convolutional code is defined as the minimum Hamming weight between any two differing paths in the state transition diagram \cite[Chapter 3]{JZ15}. Due to linearity of convolutional codes, $\dfree$ is also the minimum Hamming weight over the nonzero paths. We denote by $\Afree$ the multiplicity of paths that have Hamming weight $\dfree$. To find a good high rate code, we use $\dfree$ and $\Afree$ to select the best encoder. The BEAST algorithm described in~\cite[Chapter 10]{JZ15} is a fast method to compute $\dfree$ and $\Afree$.  The selection criterion is as follows: Keep the code with largest $\dfree$ and in case of a tie decide for the code with smaller $\Afree$.

\section{Design of Nested Convolutional Codes for PUFs}\label{sec:nestedconvforPUFs}
Algorithms~\ref{alg:designFecCode} and ~\ref{alg:designVQCode} are combined to find good nested code constructions for the coding problem described in Section~\ref{sec:problem_settingandcode}. Two TBCCs $\fecCode$ and $\quantCode$ of the same length $N$ are needed such that $\fecCode \subseteq \quantCode$. Let $\quantDim$ and $\fecDim$ denote the dimensions of $\quantCode$ and $\fecCode$, respectively, and let $\quantR = \frac{\quantDim}{N}$ and $\fecR = \frac{\fecDim}{N}$ denote their code rates. The objective is to maximize the key vs. storage rate ratio. Since $R_s =\dfrac{\fecDim}{N}$ and $R_w = \dfrac{\quantDim-\fecDim}{N}$, we have
\begin{equation}\label{eq:rsrlcomp}
    \frac{R_s}{R_w} = \frac{\fecDim}{\quantDim-\fecDim} = {\left(\frac{\quantR}{\fecR}-1\right)}^{-1}.
\end{equation}
Therefore, we maximize $\fecR$ and minimize $\quantR$ simultaneously.

To reconstruct the key $S$ of size $\fecDim$ (in bits) the code $\fecCode$ has to correct errors on the artifical BSC channel with crossover probability $\pc = q * \pA$ at a given target $\pB$. The code $\quantCode$ serves as a vector quantizer with average distortion $q$ such that \cite{OurWZTrans}
\begin{equation}\label{eq:distortionreq}
    q \leq \frac{\pc - \pA}{1-2\pA}.
\end{equation}
The design procedure is then as follows:
\begin{enumerate}
    \item Choose $m$ and $n$ to design a TBCC of rate $\fecR=\dfrac{1}{n}$ by using Algorithm~\ref{alg:designFecCode}.
    \item Obtain the corresponding value of $\pc$ where the code achieves the target block error probability $\pB$ by Monte Carlo simulations.
    \item Construct code $\quantCode$ from $\fecCode$ by using Algorithm~\ref{alg:designVQCode} such that (\ref{eq:distortionreq}) is satisfied.
\end{enumerate}
The last step in this procedure is executed by applying Algorithm~\ref{alg:designVQCode} incrementally as follows:
\begin{enumerate}
    \item Initialization: Start constructing a code  $\quantCode^{(0)}$ of rate $\quantR^{(0)} = \dfrac{2}{n}$ from code $\fecCode$ (Algorithm~\ref{alg:designFecCode}).
    \item Set $i \leftarrow 1$.
    \item Construct a code $\quantCode^{(i)}$ of rate $\quantR^{(i)} = \dfrac{i+2}{n}$ from code $\quantCode^{(i-1)}$ (Algorithm~\ref{alg:designVQCode}).
    \item If the average distortion achieved by the code $\quantCode^{(i)}$ satisfies the constraint given in (\ref{eq:distortionreq}), stop; else increment $i\leftarrow i+1$ and go to step (3).
\end{enumerate}
The final code $\quantCode$ is the code in the last iteration.
To obtain code rates in between those steps we randomly freeze inputs of the encoder in a time-variant manner as described in Section~\ref{sec:nestedconv}. Since in each iteration the code is optimized for the minimum distance of the code, we can only freeze inputs on the last added input. This way we guarantee to preserve the minimum distance of the code for the next iteration due to $\quantCode^{(i-1)} \subseteq \quantCode^{(i)}$.


\section{Estimated Decoding Complexity}\label{sec:decCompEst}
We compare the decoding complexities of TBCCs and PCs. Since the real complexity of decoding depends on the hardware implementation, we only estimate the complexity for both code classes by using standard decoding algorithms. 

The WAVA algorithm performs standard Viterbi decoding on the tailbiting trellis of the TBCC in a circular fashion. That means the decoder runs over the trellis several times and at each iteration the probabilities of the starting states of the trellis are updated according to the probabilities of the ending states of the previous iteration. Therefore, the WAVA algorithms scales with the complexity of a standard Viterbi decoder times the number of iterations. For simplicity, we consider the worst case complexity and hence let $V$ denote the number of maximum iterations of the WAVA decoder.

According to~\cite{sido19}, let $\kappa$ be the complexity of a standard Viterbi decoder with indices
\begin{itemize}
    \item \emph{F} for Forney trellis,
    \item \emph{P} for precomputation,
    \item \emph{M} for merged or minimal trellis.
\end{itemize}
We have for the total of number $\frac{N}{n}$ of trellis sections
\begin{align}
\kappa_{F} &\propto N \cdot 2^{k+m}\\
\kappa_{P} &\propto \frac{N}{n} \left(2^{k+m} + 2^{n}\right)\\
\kappa_{M} &\propto N \cdot 2^{\min\{k,n-k\}+m}.
\end{align}
By scaling these complexities with the maximum number of WAVA iterations $V$ we obtain the desired complexities of decoding a TBCC. For decoding on the Forney trellis, we can reuse the branch metrics computed in the first WAVA iteration in the following $V-1$ iterations and; therefore, we obtain
\begin{align}
    \kappa_{F}^\text{WAVA} &\propto (n+V-1) \frac{N}{n} 2^{k+m}\\
    \kappa_{P}^\text{WAVA} &\propto  \frac{N}{n} \left( V \cdot 2^{k+m} + 2^{n}\right)\\
    \kappa_{M}^\text{WAVA} &\propto V N \cdot 2^{\min\{k,n-k\}+m}.
\end{align}
Overall we have that the complexity $\kappa^\text{WAVA}$ of decoding a TBCC is 
\begin{equation}
 \kappa^\text{WAVA} \propto \min \left\{ \kappa_{F}^\text{WAVA}, \kappa_{P}^\text{WAVA}, \kappa_{M}^\text{WAVA} \right\}.   
\end{equation}

For error correction and vector quantization, we obtain different complexities since we have different values for $k$. For the error correcting code we have $k=\feck=1$ and for the vector quantizer code we have $k=\quantk$, where $\quantk$ is the largest value needed to achieve a rate of $\quantR$ such that $\quantk=\left\lceil n \quantR \right\rceil$.
The complexity of the vector quantizer can be reduced by considering decoding over the trellis with the time-variant frozen input bit values, since all branches that do not correspond to the frozen input bit value can be removed. For simplicity, we will only consider the complexity over the time-invariant trellis.

For the PCs under successive cancellation list (SCL) decoding \cite{talvardyscl} with a list size $L$, we have a complexity proportional to $L  N  \log_2 N$. This complexity is independent of the code rate and thus applies to $\fecCode$ and $\quantCode$. All decoding complexities are summarized in Table~\ref{tab:complexities}. 

Note that for the Viterbi decoder parallelization up to a factor of $2^m$ can be easily achieved since all state nodes in a trellis section can be processed independently. For the SCL decoding of PCs, parallelization cannot be achieved without changing the decoder's error correction performance since each decoded bit sequentially depends on the previously decoded ones.

\begin{table}[t]
\renewcommand{\arraystretch}{2.2} 
  \caption{Complexities of the error correcting code $\fecCode$ and vector quantizer code $\quantCode$ for PCs and TBCCs.}
  \label{tab:complexities}
  \begin{tabular}{l|c|c}
    \toprule
    Code class & Complexity of $\fecCode$ & Complexity of $\quantCode$ \\
    \midrule
    TBCC $\kappa_{F}^\text{WAVA}$ & $ \propto (n+V-1) \frac{N}{n} 2^{m}$ & $\propto (n+V-1) \frac{N}{n} 2^{\quantk+m}$ \\
    TBCC $\kappa_{P}^\text{WAVA}$ & $\propto \frac{N}{n} \left( V \cdot 2^{m} + 2^{n}\right)$ & $\propto \frac{N}{n} \left( V \cdot 2^{\quantk+m} + 2^{n}\right)$ \\
    TBCC $\kappa_{M}^\text{WAVA}$ & $\propto V N \cdot 2^{1+m}$ & $\propto V N \cdot 2^{\min\{\quantk,n-\quantk\}+m}$ \\\hline
    PC & $ \propto L  N  \log_2 N$ & $ \propto L  N  \log_2 N $ \\
    \bottomrule
\end{tabular}
\end{table}

\section{Performance Evaluations for PUFs}\label{sec:results}

\begin{table*}[t]
\renewcommand{\arraystretch}{2.8} 
  \caption{Parameters of the designed codes for $\fecDim=128$ bits, $\pA=0.0149$ and $\pB\leq 10^{-6}$ and complexities for $\fecCode$ and $\quantCode$, respectively. For the TBCCs also the type of complexity ($\kappa_{F}^\text{WAVA}$,$\kappa_{M}^\text{WAVA}$ or $\kappa_{P}^\text{WAVA}$) which is minimal is given. $\left\lceil\log_2|\mathcal{W}|\right\rceil$ is the amount of helper data in bits.}
  \label{tab:res}
  \begin{tabular}{l|c|c|c|c|c|c|c|c|c|c}
    \toprule
    Code & $m$ & $\fecR$ & $\pc$ & $\bar{q}$ & $\quantR$ & $R_w$ & $\left\lceil\log_2|\mathcal{W}|\right\rceil$ & $\displaystyle \frac{R_s}{R_w}$ & Complexity $\fecCode$ & Complexity $\quantCode$ \\
    \midrule
    \ref{plot:tbccm11n384}~$\quad$~TBCC & $11$ & $\frac{1}{3}$ & $0.0545$ & $0.0408$ & $0.8047$ & $0.4714$ & $181$ &  $0.7072$& $\kappa_{P}^\text{WAVA}\propto 2^{21.00}$& $\kappa_{M}^\text{WAVA}\propto 2^{21.58}$\\
    \ref{plot:tbccm8n384}~$\quad$~TBCC & $8$ & $\frac{1}{3}$ & $0.0365$ & $0.0223$ & $0.8906$ & $0.5573$ & $214$ & $0.5981$& $\kappa_{P}^\text{WAVA}\propto 2^{18.00}$ & $\kappa_{M}^\text{WAVA}\propto 2^{18.58}$\\   
    \ref{plot:tbccm11n512}~$\quad$~TBCC & $11$ & $\frac{1}{4}$ & $0.0837$ & $0.0709$ & $0.6680$ & $0.4180$ & $214$ & $0.5981$& $\kappa_{P}^\text{WAVA}\propto 2^{21.00}$ & $\kappa_{M}^\text{WAVA}\propto 2^{23.00}$\\
    \ref{plot:tbccm8n512}~$\quad$~TBCC & $8$ & $\frac{1}{4}$ & $0.0640$ & $0.0507$ & $0.7441$ & $0.4941$ & $253$ & $0.5059$& $\kappa_{P}^\text{WAVA}\propto 2^{18.01}$  & $\kappa_{M}^\text{WAVA}\propto 2^{20.00}$\\   
               \ref{plot:pc512}~$\quad$~PC & - & $\frac{1}{4}$ & $0.0778$ & $0.0648$ & $0.6875$ &  $0.4375$ & $224$ & $0.5714$& $\propto 2^{15.17}$ & $\propto 2^{15.17}$ \\
               \ref{plot:pc1024}~$\quad$~PC & - & $\frac{1}{8}$ & $0.1819$ & $0.1721$ & $0.3584$ &  $0.2333$  & $239$ & $0.5358$ & $ \propto 2^{16.32}$ & $ \propto 2^{16.32}$\\               
  \bottomrule
\end{tabular}
\end{table*}

In this section, the performance of TBCCs designed by the proposed procedure for the PUF setting is presented. We consider PUF devices with $\pA = 0.0149$, target block error probability $\pB = 10^{-6}$ and a key size of $\fecDim = 128$ bits. These values correspond to the best RO PUF designs in the literature \cite{OnurnewICASSPPaper}. We construct TBCCs with rates $\fecR = \frac{1}{3}$ and $\fecR = \frac{1}{4}$, and with memories $m=8$ and $m=11$. As a reference, we also give a PC construction using the approach described in~\cite{OurWZTrans}. Without puncturing we can only provide a PC construction for the case of $\fecR=\frac{1}{4}$, since for a key size of $\fecDim=128$ and code rate $\fecR=\frac{1}{3}$ we would have $N=384$ which is not a power of two. All simulations for the PCs are performed by using SCL decoding with a list size of $L=8$. We also compute the results for the rate $\frac{1}{8}$ PC presented in~\cite{OurWZTrans} but now for $\pA = 0.0149$ and give the resulting key vs. storage rate ratio $\dfrac{R_s}{R_w}$. All simulations for the TBCCs are performed by using the WAVA algorithm with a maximum of $V=4$ iterations. The final results of all discussed codes are given in Table~\ref{tab:res}.

\subsection{Error Correction Performance}
\begin{figure}
	\centering
    \begin{tikzpicture}
    
    \begin{semilogyaxis}[
    yticklabel style={/pgf/number format/fixed},
    xticklabel style={/pgf/number format/fixed},
    width=\linewidth,
    height=9.1cm,
    ymax = 1.0e-4,
    ymin = 1.0e-7,
    xmin = 0.04,
    xmax = 0.2,
    ylabel={Block Error Probability $\pB$},
    xlabel={Crossover Probability $\pc$},
    legend cell align=left,
    grid=both,
    mark options=solid,
    legend columns=1,
    minor grid style={dotted},
    major grid style={dotted,black},
    legend style={at={(0.99,0.90)},anchor=north east,thick,font={\normalsize }},
    ]
    
    \addplot[black,solid, very thick] table [x=p, y=mc, col sep=comma] {
    	p,na,mc,rcu
    	0.123,3.9505e-08,4.9021e-10,1.6774e-09
    	0.12754,1.796e-07,3.8633e-09,1.3412e-08
    	0.13208,7.3697e-07,2.6152e-08,8.0467e-08
    	0.13662,2.7491e-06,1.5356e-07,4.2915e-07
    	0.14115,9.3804e-06,7.8898e-07,2.1458e-06
    	0.14569,2.9441e-05,3.5758e-06,8.5831e-06
    	0.15023,8.5408e-05,1.44e-05,3.4332e-05
    	0.15477,0.00022999,5.1872e-05,0.00010986
    	0.15931,0.00057712,0.00016818,0.00032959
    	0.16385,0.001354,0.00049355,0.00087891
    	0.16838,0.0029792,0.0013181,0.002417
    	0.17292,0.0061641,0.0032194,0.0052734
    	0.17746,0.012023,0.0072251,0.012305
    	0.182,0.022154,0.014966,0.022852
    	0.18654,0.038647,0.028733,0.042188
    	0.19108,0.063947,0.051344,0.070313
    	0.19562,0.10055,0.085743,0.1125
    	0.20015,0.1505,0.13435,0.17578
    	0.20469,0.21483,0.19831,0.25313
    	0.20923,0.29301,0.27688,0.3375
    	0.21377,0.38264,0.36718,0.43594
    	0.21831,0.47954,0.46449,0.53438
    	0.22285,0.57823,0.56301,0.61875
    	0.22738,0.67283,0.65687,0.71719
    	0.23192,0.75806,0.74118,0.7875
    	0.23646,0.83013,0.81269,0.85078
    	0.241,0.88721,0.87005,0.89824
    	0.24554,0.9,0.9,0.9
    	0.25008,0.9,0.9,0.9
    	0.25462,0.9,0.9,0.9
    	0.25915,0.9,0.9,0.9
    	0.26369,0.9,0.9,0.9
    	0.26823,0.9,0.9,0.9
    	0.27277,0.9,0.9,0.9
    	0.27731,0.9,0.9,0.9
    	0.28185,0.9,0.9,0.9
    	0.28638,0.9,0.9,0.9
    	0.29092,0.9,0.9,0.9
    	0.29546,0.9,0.9,0.9
    	0.3,0.9,0.9,0.9
    };
    \addlegendentry{MC};
    
    \addplot[black,dotted,very thick] table [x=p, y=rcu, col sep=comma] {
    	p,na,mc,rcu
    	0.123,3.9505e-08,4.9021e-10,1.6774e-09
    	0.12754,1.796e-07,3.8633e-09,1.3412e-08
    	0.13208,7.3697e-07,2.6152e-08,8.0467e-08
    	0.13662,2.7491e-06,1.5356e-07,4.2915e-07
    	0.14115,9.3804e-06,7.8898e-07,2.1458e-06
    	0.14569,2.9441e-05,3.5758e-06,8.5831e-06
    	0.15023,8.5408e-05,1.44e-05,3.4332e-05
    	0.15477,0.00022999,5.1872e-05,0.00010986
    	0.15931,0.00057712,0.00016818,0.00032959
    	0.16385,0.001354,0.00049355,0.00087891
    	0.16838,0.0029792,0.0013181,0.002417
    	0.17292,0.0061641,0.0032194,0.0052734
    	0.17746,0.012023,0.0072251,0.012305
    	0.182,0.022154,0.014966,0.022852
    	0.18654,0.038647,0.028733,0.042188
    	0.19108,0.063947,0.051344,0.070313
    	0.19562,0.10055,0.085743,0.1125
    	0.20015,0.1505,0.13435,0.17578
    	0.20469,0.21483,0.19831,0.25313
    	0.20923,0.29301,0.27688,0.3375
    	0.21377,0.38264,0.36718,0.43594
    	0.21831,0.47954,0.46449,0.53438
    	0.22285,0.57823,0.56301,0.61875
    	0.22738,0.67283,0.65687,0.71719
    	0.23192,0.75806,0.74118,0.7875
    	0.23646,0.83013,0.81269,0.85078
    	0.241,0.88721,0.87005,0.89824
    	0.24554,0.9,0.9,0.9
    	0.25008,0.9,0.9,0.9
    	0.25462,0.9,0.9,0.9
    	0.25915,0.9,0.9,0.9
    	0.26369,0.9,0.9,0.9
    	0.26823,0.9,0.9,0.9
    	0.27277,0.9,0.9,0.9
    	0.27731,0.9,0.9,0.9
    	0.28185,0.9,0.9,0.9
    	0.28638,0.9,0.9,0.9
    	0.29092,0.9,0.9,0.9
    	0.29546,0.9,0.9,0.9
    	0.3,0.9,0.9,0.9
    };
    \addlegendentry{RCU};
    
    \addplot[mark=*,red,dashed,very thick] table [x=p, y=fer, col sep=comma] {
    	p,fer,ber
    	0.12,0.0005563445236666032,7.343198688198176e-05
    	0.11333333333333333,0.00029819654092012534,3.704488277935976e-05
    	0.10666666666666666,0.0001465765988366006,1.657166234614692e-05
    	0.09999999999999999,4.5735820752171306e-05,5.711873149741037e-06
    	0.09333333333333332,1.7778274440682787e-05,1.9425145845344253e-06
    	0.08666666666666667,5.051908722979234e-06,4.933504612284409e-07
    	0.07999999999999999,2.019460036198533e-06,1.9991752813706455e-07
    	0.07333333333333333,5.1848498982251e-07,4.7971434884247855e-08
    	0.06666666666666667,1.4937669958007968e-07,1.5171071051101845e-08
    };
    \addlegendentry{PC, $L=8$ (simul.)};
    
    \addplot[blue,solid, very thick] table [x=p, y=fer, col sep=comma] {
    	p,fer
    	0.02,3.6246230411083963e-16
    	0.02947368421052632,1.3101804560987223e-13
    	0.03894736842105263,9.259738030621198e-12
    	0.04842105263157895,2.623213249400081e-10
    	0.05789473684210526,4.139098144951934e-09
    	0.06736842105263158,4.3551962759659334e-08
    	0.0768421052631579,3.4154682179881167e-07
    	0.08631578947368422,2.15423496738243e-06
    	0.09578947368421054,1.1585241377283395e-05
    	0.10526315789473685,5.603169606139737e-05
    	0.11473684210526316,0.0002603730609486749
    	0.12421052631578948,0.001384640727835957
    	0.1336842105263158,0.04538403676216494
    	0.14315789473684212,1.0
    	0.15263157894736842,1.0
    	0.16210526315789472,1.0
    	0.17157894736842105,1.0
    	0.18105263157894738,1.0
    	0.19052631578947368,1.0
    	0.2,1.0
    };
    \addlegendentry{TBCC, $m=11$ (UB)};
    
    \addplot[mark=+,blue,dashed,very thick] table [x=p, y=fer, col sep=comma] {
    	p,fer,ber
    	0.12,0.0002420302883618007,9.96754200954291e-06
    	0.11333333333333333,0.00010218910970657857,3.689528681928367e-06
    	0.10666666666666666,4.466447725652915e-05,1.4755229093674809e-06
    	0.09999999999999999,1.5197002543218376e-05,4.503129659848749e-07
    	0.09333333333333332,5.798643672393697e-06,1.653519484706015e-07
    	0.08666666666666667,1.5808306309692377e-06,4.146151766492978e-08
    	0.07999999999999999,5.908635944123211e-07,1.500239595187534e-08
    };
    \addlegendentry{TBCC, $m=11$ (simul.)};
    
    \addplot[cyan,solid, very thick] table [x=p, y=fer, col sep=comma] {
    	p,fer
    	0.02,8.629064597905065e-13
    	0.02947368421052632,9.946834638984518e-11
    	0.03894736842105263,3.088916669084225e-09
    	0.04842105263157895,4.601418395905571e-08
    	0.05789473684210526,4.2861834371059193e-07
    	0.06736842105263158,2.8875467929488546e-06
    	0.0768421052631579,1.5398024288007502e-05
    	0.08631578947368422,6.917243093145192e-05
    	0.09578947368421054,0.00027485832917826766
    	0.10526315789473685,0.0010118990932230082
    	0.11473684210526316,0.0036698956171199995
    	0.12421052631578948,0.015054577956118809
    	0.1336842105263158,0.14690557110756639
    	0.14315789473684212,1.0
    	0.15263157894736842,1.0
    	0.16210526315789472,1.0
    	0.17157894736842105,1.0
    	0.18105263157894738,1.0
    	0.19052631578947368,1.0
    	0.2,1.0
    };
    \addlegendentry{TBCC, $m=8$ (UB)};
    
    \addplot[mark=diamond*, cyan, dashed, very thick] table [x=p, y=fer, col sep=comma] {
    	p,fer,ber
    	0.12,0.002557320084723429,8.853317683575563e-05
    	0.11333333333333333,0.0012493731138425261,4.206736056732098e-05
    	0.10666666666666666,0.0006556828169249871,2.0562747915533173e-05
    	0.09999999999999999,0.00030166170672063774,7.655200669218312e-06
    	0.09333333333333332,0.0001266800947567109,3.273040394662787e-06
    	0.08666666666666667,5.346807287698333e-05,1.4679850365778907e-06
    	0.07999999999999999,1.8642856010969456e-05,4.5462768285678865e-07
    	0.07333333333333333,5.9179281435019795e-06,1.3804095781159752e-07
    	0.06666666666666667,1.7492940286623826e-06,3.9437209128326036e-08
    	0.06,4.2215389995504734e-07,7.763615693310467e-09
    };
    \addlegendentry{TBCC, $m=8$ (simul.)};
    
    \end{semilogyaxis}
    \end{tikzpicture}
	\caption{Error correcting performance of different codes with $\fecDim=128$ bits and $\fecR=\frac{1}{4}$ over a BSC with crossover probability $\pc$. The MC and RCU bounds for the same code parameters are given as references.}\label{fig:fecr1_4}
\end{figure}
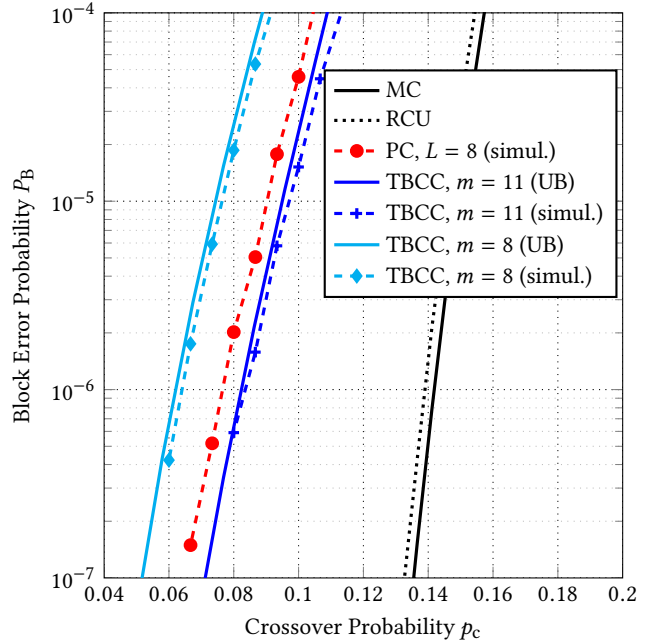

The construction of the nested code design starts with the error correcting code $\fecCode$. We design two TBCCs with $\fecR=\dfrac{1}{3}$ and $\fecR=\dfrac{1}{4}$ by using Algorithm~\ref{alg:designFecCode} with $W_\text{max}=10^4$. Results of the Monte Carlo simulations as well as the bound~\eqref{eq:ubbound} are shown in Figures~\ref{fig:fecr1_3} and ~\ref{fig:fecr1_4} for the two TBCCs.

\begin{figure}
	\centering
    \begin{tikzpicture}
    
    \begin{semilogyaxis}[
    yticklabel style={/pgf/number format/fixed},
    xticklabel style={/pgf/number format/fixed},
    width=\linewidth,
    height=9.1cm,
    ymax = 1.0e-4,
    ymin = 1.0e-7,
    xmin = 0.02,
    xmax = 0.15,
    ylabel={Block Error Probability $\pB$},
    xlabel={Crossover Probability $\pc$},
    legend cell align=left,
    grid=both,
    mark options=solid,
    legend columns=1,
    minor grid style={dotted},
    major grid style={dotted,black},
    legend style={at={(0.99,0.90)},anchor=north east,thick,font={\normalsize }},
    ]
    
    \addplot[black,solid,very thick] table [x=p, y=mc, col sep=comma] {
    	p,na,mc,rcu
    	0.075,8.4427e-10,1.1317e-11,5.3387e-11
    	0.086842,1.1944e-07,5.5001e-09,2.0118e-08
    	0.098684,5.7625e-06,7.0334e-07,1.7166e-06
    	0.11053,0.00012452,3.182e-05,6.8665e-05
    	0.12237,0.0014351,0.00062646,0.0010986
    	0.13421,0.0099246,0.0062479,0.010547
    	0.14605,0.044734,0.035483,0.049219
    	0.15789,0.13977,0.12617,0.16875
    	0.16974,0.31819,0.30486,0.36563
    	0.18158,0.55282,0.54004,0.59063
    	0.19342,0.77036,0.75529,0.80156
    	0.20526,0.9,0.8969,0.9
    	0.21711,0.9,0.9,0.9
    	0.22895,0.9,0.9,0.9
    	0.24079,0.9,0.9,0.9
    	0.25263,0.9,0.9,0.9
    	0.26447,0.9,0.9,0.9
    	0.27632,0.9,0.9,0.9
    	0.28816,0.9,0.9,0.9
    	0.3,0.9,0.9,0.9   
};
    \addlegendentry{MC};
    
    \addplot[black,dotted,very thick] table [x=p, y=rcu, col sep=comma] {
    	p,na,mc,rcu
    	0.075,8.4427e-10,1.1317e-11,5.3387e-11
    	0.086842,1.1944e-07,5.5001e-09,2.0118e-08
    	0.098684,5.7625e-06,7.0334e-07,1.7166e-06
    	0.11053,0.00012452,3.182e-05,6.8665e-05
    	0.12237,0.0014351,0.00062646,0.0010986
    	0.13421,0.0099246,0.0062479,0.010547
    	0.14605,0.044734,0.035483,0.049219
    	0.15789,0.13977,0.12617,0.16875
    	0.16974,0.31819,0.30486,0.36563
    	0.18158,0.55282,0.54004,0.59063
    	0.19342,0.77036,0.75529,0.80156
    	0.20526,0.9,0.8969,0.9
    	0.21711,0.9,0.9,0.9
    	0.22895,0.9,0.9,0.9
    	0.24079,0.9,0.9,0.9
    	0.25263,0.9,0.9,0.9
    	0.26447,0.9,0.9,0.9
    	0.27632,0.9,0.9,0.9
    	0.28816,0.9,0.9,0.9
    	0.3,0.9,0.9,0.9
    };
    \addlegendentry{RCU};

    \addplot[blue,solid,  very thick] table [x=p, y=fer, col sep=comma] {
    	p,fer
    	0.02,1.2260398848685262e-11
    	0.02947368421052632,1.2823563892834159e-09
    	0.03894736842105263,3.6792325511119503e-08
    	0.04842105263157895,5.187590489919527e-07
    	0.05789473684210526,4.727951567806361e-06
    	0.06736842105263158,3.2740471849697696e-05
    	0.0768421052631579,0.00019336732762325232
    	0.08631578947368422,0.0010453695658573792
    	0.09578947368421054,0.005101589407361638
    	0.10526315789473685,0.02083760500566494
    	0.11473684210526316,0.06710512302726021
    	0.12421052631578948,0.16771858283570149
    	0.1336842105263158,0.33032648434610423
    	0.14315789473684212,0.5289591123507041
    	0.15263157894736842,0.7160846365452783
    	0.16210526315789472,0.8547597355012494
    	0.17157894736842105,0.9370864175337702
    	0.18105263157894738,0.9768769407485606
    	0.19052631578947368,0.9927577340797676
    	0.2,0.9980565959862204
    };
    \addlegendentry{TBCC, $m=11$ (UB)};
    
    \addplot[mark=+,blue,dashed,very thick] table [x=p, y=fer, col sep=comma] {
    	p,fer,ber
    	0.08,0.00016381223841233178,8.199753228563706e-06
    	0.07333333333333333,4.8974046204214075e-05,2.276527929024014e-06
    	0.06666666666666667,1.832715969868055e-05,8.222676561238371e-07
    	0.06,4.060520786153949e-06,1.6926389661032365e-07
    	0.05333333333333333,1.0768147659240059e-06,4.110163503861719e-08
    	0.04666666666666666,2.1429507761105088e-07,7.930327707645797e-09
    };
    \addlegendentry{TBCC, $m=11$ (simul.)};
    
    \addplot[cyan,solid,very thick] table [x=p, y=fer, col sep=comma] {
    	p,fer
    	0.02,3.756802496156066e-09
    	0.02947368421052632,1.5751694448963187e-07
    	0.03894736842105263,2.3883132590634597e-06
    	0.04842105263157895,2.0635605432182583e-05
    	0.05789473684210526,0.00012626425734170676
    	0.06736842105263158,0.0006246495498973362
    	0.0768421052631579,0.0027964943364586115
    	0.08631578947368422,0.013649495615287994
    	0.09578947368421054,0.19438811525962574
    	0.10526315789473685,1.0
    	0.11473684210526316,1.0
    	0.12421052631578948,1.0
    	0.1336842105263158,1.0
    	0.14315789473684212,1.0
    	0.15263157894736842,1.0
    	0.16210526315789472,1.0
    	0.17157894736842105,1.0
    	0.18105263157894738,1.0
    	0.19052631578947368,1.0
    	0.2,1.0
    };
    \addlegendentry{TBCC, $m=8$ (UB)};
    
    \addplot[mark=square*,cyan,dashed,very thick] table [x=p, y=fer, col sep=comma] {
    	p,fer,ber
    	0.08,0.001667135144565818,6.474384759283132e-05
    	0.07333333333333333,0.0008507629906116507,3.182872103960841e-05
    	0.06666666666666667,0.00031804606931148667,1.0908162505751897e-05
    	0.06,0.000104526590239572,3.2549543395465313e-06
    	0.05333333333333333,3.181492801872536e-05,8.486347987137679e-07
    	0.04666666666666666,7.251884790756706e-06,2.1488564865467696e-07
    	0.039999999999999994,2.0820427456760247e-06,5.762796885353283e-08
    	0.03333333333333333,4.707899992716879e-07,9.195117173275154e-09
    };
    \addlegendentry{TBCC, $m=8$ (simul.)};

    \end{semilogyaxis}
    \end{tikzpicture}
	\caption{Error correcting performance of different codes with $\fecDim=128$ bits and $\fecR=\frac{1}{3}$ over a BSC with crossover probability $\pc$. The MC and RCU bounds for the same code parameters are given as references.}\label{fig:fecr1_3}
\end{figure}
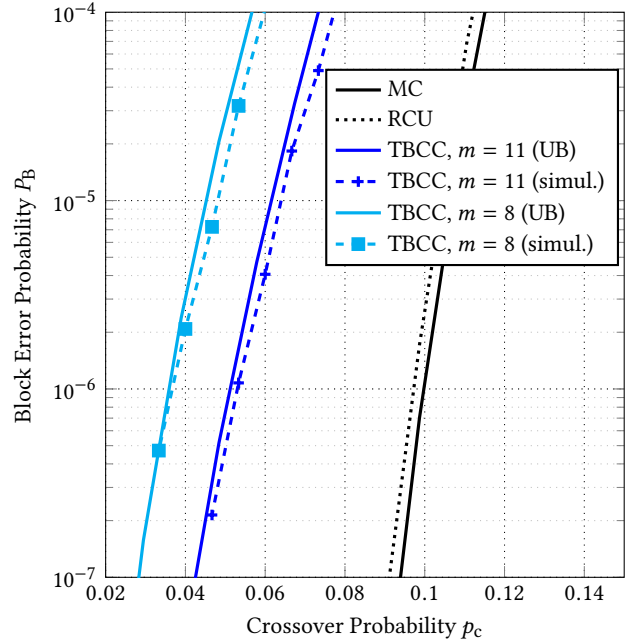

To bound the code performance on a BSC for a given block length and code rate we use two finite length bounds, namely the \textit{meta converse (MC)} and the \textit{random coding union (RCU)} bound from~\cite{poly10}. The MC gives a lower bound and the RCU an upper bound on the block error probability. For $\fecR=\frac{1}{4}$, we observe that the TBCC with $m=11$ outperforms the PC, whereas the TBCC with $m=8$ performs worse. We also observe that for all considered codes there is still a gap to the finite length bounds.

\begin{figure}
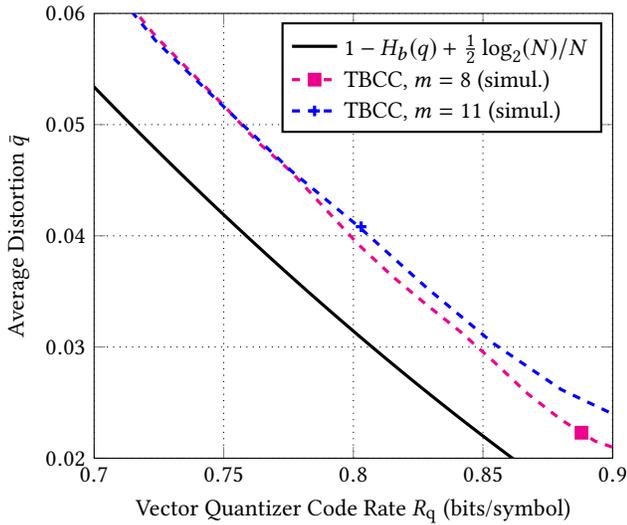

	\centering

	\caption{Code rate of the vector quantizer code $\quantCode$ vs. average distortion $\bar{q}$ for $N=384$ bits and $P_B \leq 10^{-6}$.}\label{fig:distortionplot384}
\end{figure}

\subsection{Vector Quantization Performance}\label{subsec:vectorQPerformance}
\begin{figure}
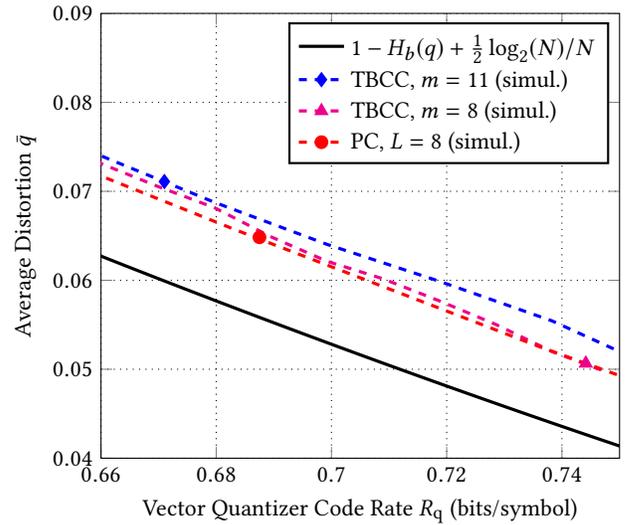

\newpage
	\centering

	\caption{Code rate of the vector quantizer code $\quantCode$ vs. average distortion $\bar{q}$ for $N=512$ bits and $P_B\leq 10^{-6}$.}\label{fig:distortionplot512}
\end{figure}

Using the approach described in Section~\ref{sec:nestedconv} and setting $W_\text{max}=10^4$, we construct high rate codes to be used as a vector quantizer. Using Monte Carlo simulations, we plot the rate of these codes $\quantR$ vs. the measured average distortion $\bar{q}$ in Figures~\ref{fig:distortionplot384} and ~\ref{fig:distortionplot512} for $N=384$, corresponding to $\fecR=\frac{1}{3}$, and for $N=512$, corresponding to $\fecR=\frac{1}{4}$, respectively.

\begin{figure*}[t]
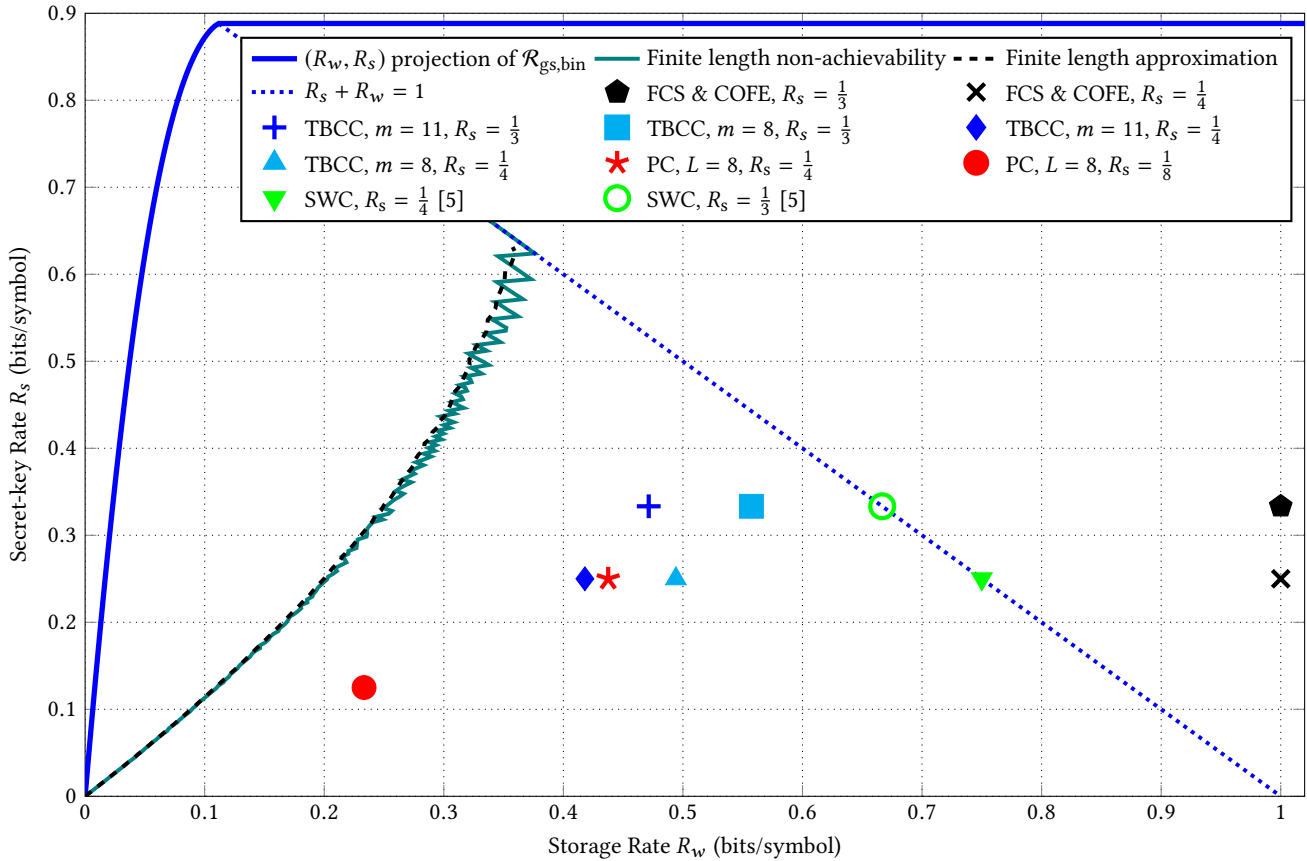

	\centering
	%
	\caption{Storage-key rates for the GS model with $p_A=0.0149$. The $(0.1118,0.8882)$ bits/symbol point is the best possible point achieved by SW-coding (SWC) constructions such as polar codes (PCs) in \cite{IgnaPolar}, which lies on the dashed line representing $R_w+R_s = H(X)$. The PCs are designed by applying the design procedure proposed in \cite{OurWZTrans} for WZ-coding with the SCL decoder with list size of $L$. The block-error probability satisfies $P_B \leq 10^{-6}$ and $\fecDim=128$ bits for all codes. The finite length non-achievability bound and its approximation for $\fecDim=128$ bits is depicted as well.}
	\label{fig:rateregionplot}
\end{figure*}

We plot the approximate bound on the rate achieved for a given distortion from \cite{kostina}. The approximated rate for block length $N$ is
\begin{equation}\label{eq:kostinabound}
    \quantR^{(\text{approx})} \; \eqdef 1-H_b(q)+\frac{\log_2(N)}{2N}+O\left(\frac{1}{N}\right)
\end{equation}
where $O(\cdot)$ denotes the big $O$ notation. This approximation does not consider the effect of the constraint that the error correcting code designed in the previous subsection has to be a subcode of the vector quantizer of rate $\quantR$. Therefore, this bound only gives an approximate achievable bound on the rate of the high-rate code that is used as a vector quantizer without having any constraint. The bound is plotted by neglecting the $\displaystyle O\left(\frac{1}{N}\right)$ term.

Using \eqref{eq:distortionreq}, we obtain the target distortion for the code to be designed, which allows to find a lower bound on the required rate $\quantR$ of the vector quantizer. The results are shown in Table~\ref{tab:res}. Observe that vector quantization performance of all codes is similar. Therefore, the code that has the best error correction performance yields the smallest rate for vector quantization, which corresponds to the smallest amount of helper data.

\subsection{Overall Performance}


Combining the results of the error correction and the vector quantizer performance, we can evaluate the key vs. storage rate ratio by using~\eqref{eq:rsrlcomp}. The intermediate and final results are listed in Table~\ref{tab:res}, and the achieved $(R_w, R_s)$ tuples for all mentioned codes are depicted in Figure~\ref{fig:rateregionplot}. 

For the nested WZ-coding construction, where we have a vector quantizer and an error correcting code, we plot a finite length non-achievability (converse) bound. For a fixed key size of $\fecDim=128$ bits and $\pA=0.0149$, we evaluate the MC non-achievability bound for the error correcting code and combine this bound using \eqref{eq:distortionreq} with the non-achievability bound from \cite[(2.186)]{kostina} for the vector quantizer code. A slightly tighter version of the non-achievability bound for the vector quantizer code can be found in \cite{RoyRDConverse}. To achieve a distortion of $q$, any vector quantizer code of blocklength $N$ must satisfy \cite[(2.186)]{kostina}
\begin{equation}\label{eq:kostinaconverse}
    \sum_{j=0}^{\left\lfloor Nq \right\rfloor} \binom{N}{j} \geq 2^{N\left(1-R_q\right)}.
\end{equation}
Similar to the achievability bound discussed in Section~\ref{subsec:vectorQPerformance}, \eqref{eq:kostinabound} is used to approximate also the non-achievability bound in \eqref{eq:kostinaconverse}. The combination of the MC bound and the converse bound for the vector quantizer performance establishes a non-achievability bound on the best rate tuples that can be achieved for given parameters by our WZ-coding construction. In Figure~\ref{fig:rateregionplot}, we plot this non-achievability bound using \eqref{eq:kostinaconverse} and its approximation using \eqref{eq:kostinabound}. Note that the zigzag behaviour of the bound in \eqref{eq:kostinaconverse} is due to the floor function. We observe a gap between these bounds and achieved rate tuples by the designed codes. 

The FCS and COFE have the key vs. storage rate ratio of
\begin{align}
\frac{R_s}{R_w}=R_s    
\end{align}
as the storage rate is $1$ bit/symbol for these constructions. The SW coding constructions such as the syndrome coding method proposed in \cite{IgnaPolar} achieve the ratio 
\begin{align}
\frac{R_s}{R_w}=\frac{R_s}{1-R_s}    
\end{align}
which improves on the FCS and COFE. WZ coding constructions with nested PC we constructed for $p_A = 0.0149$ based on the design procedure given in \cite{OurWZTrans} achieves even larger ratios. The largest key vs. storage rate ratio is achieved by the TBCC with $\fecR=\frac{1}{3}$ and $m=11$ such that $\dfrac{R_s}{R_w}=0.7072$. These results suggest that increasing the code rate $\fecR$ and the memory size of TBCCs allows a larger key vs. storage rate ratio.

\section{Conclusion}\label{sec:conclusion}
We proposed a nested convolutional code construction, which might be useful for various achievability schemes. For the key agreement problem with PUFs, we proposed a design procedure for the nested code construction using TBCCs to obtain good reliability, secrecy, privacy, storage, and cost performance jointly. We implemented nested convolutional codes for practical source and channel parameters to illustrate the gains in terms of the key vs. storage rate ratio as compared to previous code designs. We observe that one variant of nested convolutional codes achieves a higher rate ratio than all other code designs in the literature but it may have a high hardware cost. Another variant of nested convolutional codes with low complexity is illustrated to perform similarly to the best previous codes in the literature. We also computed known finite-length bounds for our code construction to show the gaps between the performance of the designed codes and these bounds. 


\begin{acks}
This work was performed while O. G{\"u}nl{\"u} was with the Chair of Communications Engineering, Technical University of Munich. O. G\"unl\"u was supported by the German Federal Ministry of Education and Research (BMBF) within the national initiative for ``Post Shannon Communication (NewCom)'' under the Grant 16KIS1004, and by the German Research Foundation (DFG) under grant KR 3517/9-1. V. Sidorenko is on leave from the Institute for Information Transmission Problems, Russian Academy of Sciences. His work was supported by the European Research Council (ERC) under the European Union’s Horizon 2020 research and innovation programme (grant agreement No 801434) and by the Chair of Communications Engineering at the Technical University of Munich. The work of G. Kramer was supported by an Alexander von Humboldt Professorship endowed by the BMBF.
\end{acks}


\bibliographystyle{ACM-Reference-Format}
\bibliography{sample-base}


\end{document}